\documentclass[aps,prl,twocolumn,groupedaddress,showpacs]{revtex4}

\usepackage{graphicx}
\usepackage{dcolumn}
\usepackage{bm}

\begin{document}

\title{Two-dimensional Bose-Einstein condensate in an optical surface trap}

\author{D. Rychtarik}
\author{B. Engeser}
\author{H.-C. N\"agerl}
\author{R. Grimm}

\affiliation{Institut f\"ur Experimentalphysik, Universit\"at
Innsbruck, Technikerstra{\ss}e 25, A-6020 Innsbruck, Austria}

\date{\today}

\begin{abstract}
We report on the creation of a two-dimensional Bose-Einstein
condensate of cesium atoms in a gravito-optical surface trap. The
condensate is produced a few $\mu$m above a dielectric surface on
an evanescent-wave atom mirror. After evaporative cooling by
all-optical means, expansion measurements for the tightly confined
vertical motion show energies well below the vibrational energy
quantum. The presence of a condensate is observed in two
independent ways by a magnetically induced collapse at negative
scattering length and by measurements of the horizontal expansion.
\end{abstract}

\pacs{32.80.Pj, 03.75.-b, 34.50.-s}

\maketitle

Quantum gases in lower dimensions \cite{QGLD2003} currently
attract considerable interest as model systems to study a wide
range of phenomena related to statistical physics, condensed
matter physics, and other areas. Dimensionally reduced systems may
exhibit strikingly new properties in comparison to the
three-dimensional case. Bose-Einstein condensation (BEC) does not
occur in 1D or 2D in the case of an infinite homogeneous system
\cite{Bagnato1991a}, but it is possible in highly anisotropic
traps at finite particle number \cite{Ketterle1996a}. Only
recently, experiments have entered regimes of BEC in 1D
\cite{Gorlitz2001a,Schreck2001a} or 2D \cite{Gorlitz2001a}.

The interest in 2D BEC goes back to early experiments on
spin-polarized hydrogen on a liquid-helium surface
\cite{Walraven1991a}, for which evidence of quantum degeneracy has
been obtained \cite{Safonov1998a}. The gradual onset of coherence
and formation of quasi-condensates have already been discussed in
this context \cite{Kagan1987a}. Ultracold, laser-manipulated
atomic quantum gases offer many more intriguing features like
precise control of the external confinement, tunability of
interatomic interactions, and superb experimental access. Novel
phenomena like modified interaction properties have been predicted
for 2D quantum gases tightly confined in one direction
\cite{Petrov2000a}. The exploration of these phenomena relies on
the development of appropriate trapping schemes.

\begin{figure}
\includegraphics[width=8cm]{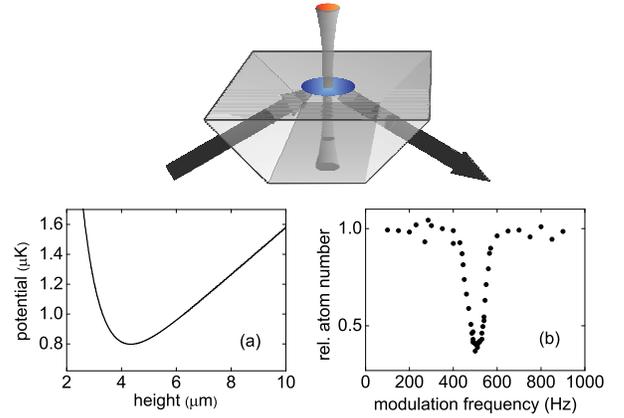}
\caption{\label{trappic} Illustration of the gravito-optical
surface trap together with (a) the calculated vertical potential
and (b) a measurement of the vertical trap frequency.}
\end{figure}


In this Letter, we report on the creation of a two-dimensional BEC
of cesium atoms by evaporative cooling in a highly anisotropic
surface trap. Our gravito-optical surface trap (GOST) is based on
an evanescent wave (EW) atom mirror in combination with a
horizontally confining optical dipole potential \cite{Grimm2000a},
see illustration in Fig.~\ref{trappic}. The repulsive EW is formed
on the surface of a prism by a blue-detuned laser beam undergoing
total internal reflection. Since gravity pushes the atoms onto the
prism the vertical motion is tightly confined in a gravito-optical
potential
\begin{equation}
U(z) = U_{0} {\rm e}^{-2z/\Lambda} + m g z\,,
\end{equation}
where $z$ is the vertical distance from the surface and $U_0$,
$m$, and $g$ denote the maximum EW potential, the atomic mass, and
the gravitational acceleration, respectively. The EW field decay
length is given by $\Lambda = \lambda/2\pi \times
(n^2\sin^2\theta-1)^{-1/2}$, where $n$ is the refractive index of
the medium. The potential minimum is located at a height $z_0 =
\frac{1}{2}\Lambda\ln(2U_0/mg\Lambda)$ and provides a trap
frequency of $\omega_z = (2g/\Lambda)^{1/2}$. The quantized nature
of the motion in such a trap and the behavior of a bosonic gas
were theoretically investigated in \cite{Wallis1992a,Wallis1996a}.

The trap loading procedure is based on methods described in our
previous work \cite{Ovchinnikov1997a,Hammes2002a}. We release
cesium atoms from a magneto-optical trap into a large-volume GOST
where horizontal confinement is realized with a blue-detuned
hollow beam (diameter 0.8\,mm, power 50\,mW, detuning 50\,GHz). In
this stage the EW is realized with a 100-mW diode laser and tuned
a few GHz above resonance with atoms in the lower hyperfine state
($F=3$). Evanescent-wave Sisyphus cooling provides a sample of
$10^7$ Cs atoms in $F=3$ at a temperature of $\sim$$10\mu$K. Then
we introduce the focussed, vertically propagating 5-W laser beam
of an Yb fibre laser (wavelength 1064\,nm, beam waist 130\,$\mu$m)
to provide a narrow dimple potential in the center of the trapped
atom cloud with a depth of $\sim$50$\mu$K and a horizontal
oscillation frequency of 130\,Hz. We wait for 1\,s until the
dimple is filled through elastic collisions with a high-density
sample of unpolarized Cs atoms in $F=3$. Finally the blue-detuned
hollow beam is turned off.

For further experiments it is crucial to strongly suppress photon
scattering and light-induced loss. Therefore the near-resonant EW
is replaced by a far-detuned EW, which is derived from the 700-mW
beam of a Ti:Sapph laser at a wavelength of 839\,nm. This EW is
produced in a nearly round spot on the surface of the fused-silica
prism ($n=1.45$) with a diameter of 500\,$\mu$m (1/${\rm e}^2$
intensity drop) and maximum potential height of $U_0 \approx k_B
\times 50$\,$\mu$K. The angle of incidence $\theta$ is set
$4.2(1.5)$\,mrad above the critical angle, which leads to
calculated values of the decay length and the vertical trap
frequency of $\Lambda =1.4(3)\mu$m and $\omega_z/2\pi=
600(50)$\,Hz. The corresponding gravito-optical potential is shown
in Fig.~\ref{trappic}(a). The potential minimum is located about
4\,$\mu$m above the surface. The trap frequency can be measured by
parametric excitation. Fig.~\ref{trappic}(b) shows a loss feature
induced by modulating the EW power at the trap frequency yielding
$\omega_z/2\pi = 550(50)\,$Hz \cite{trapfreq}.

For evaporative cooling we optically pump the atoms into the
lowest spin state $F=3$, $m_F=3$, which offers very favorable
scattering properties. In this state two-body loss is completely
suppressed and convenient magnetic tuning of the s-wave scattering
length $a$ is possible over a wide range \cite{Kerman2001a}. At a
magnetic field of $26.8\,$G where $a=440\,a_0$ ($a_0$ is Bohr's
radius) we find an optimum situation with a large elastic cross
section for evaporative cooling at moderate three-body loss
\cite{Weber2003a,Weber2003b}.
At this stage we have a sample of $10^6$ atoms at a temperature of
$9\,\mu$K with peak values for number and phase-space density of
the order of $4 \times 10^{12}$\,cm$^{-3}$ and $5 \times 10^{-4}$,
respectively.

Forced evaporative cooling is performed by ramping down the power
of the red-detuned beam by more than three orders of magnitude.
The total ramp consists of two subsequent exponential ramps. The
first ramp reduces the power from initially 5\,W down to 470\,mW
within 2\,s, the second ramp lowers the power further to a final
ramp power $P_{\rm f}$ of typically a few mW in 3.5\,s. To
counteract the density decrease when the potential is ramped down
we adiabatically reduce the waist of the red-detuned beam to
64\,$\mu$m by moving the focussing lens synchronously with the
5.5-s ramp.




\begin{figure}
\includegraphics[width=6.5cm]{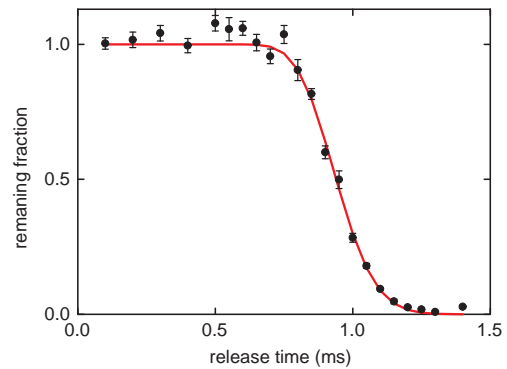}
\caption{\label{mustertemp}  Measurement of the vertical expansion
energy with 2400 atoms left after deep evaporative cooling down to
$P_{\rm f} = 1.6\,$mW. The fraction of remaining atoms is plotted
as a function of the release time. The fit of an error function to
the data (solid lines) yields an expansion energy of
$\frac{1}{2}k_B \times 16(2)\,$nK together with a release height
of $4.3(1)\,\mu$m.}
\end{figure}

A release-and-recapture method
\cite{Ovchinnikov1997a,Hammes2002a,Hammes2003a} is applied to
observe the vertical expansion of the sample and thus to measure
the vertical expansion energy $E_{\rm exp}$. We turn off the EW
for a short variable release time. Without EW the sample drops
onto the surface while undergoing a vertical expansion. Those
atoms that hit the room-temperature surface get lost. After the
short release time the EW is turned on again to prevent the
remaining atoms from hitting the surface, and their number is
measured after recapture into the MOT by taking a fluorescence
image. Corresponding measurements without release are used to
normalize the data and to determine the remaining fraction as a
function of the release time. In the classical regime of a thermal
gas with negligible effect of the quantized vertical motion,
$E_{\rm exp}$ is directly related to the temperature $T$ by
$E_{\rm exp} = \frac{1}{2}k_B T$, where $k_B$ is Boltzmann's
constant. This classical regime ranges down to temperatures of
$\sim$100\,nK.

Fig.~\ref{mustertemp} shows a measurement taken with 2400 atoms
remaining at $P_{\rm f}=1.6$\,mW, where the horizontal trap
frequency is 10\,Hz. We will show later that this sample is Bose
condensed. The position of the sharp edge corresponds to the
release height and the steepness is related to the energy spread.
A corresponding fit of an error function to the data yields a
release height of $z_0 = 4.3(1)\,\mu$m and an expansion energy
$E_{\rm exp} = \frac{1}{2}k_B \times 16(2)$\,nK.

\begin{figure}
\includegraphics[width=7.5cm]{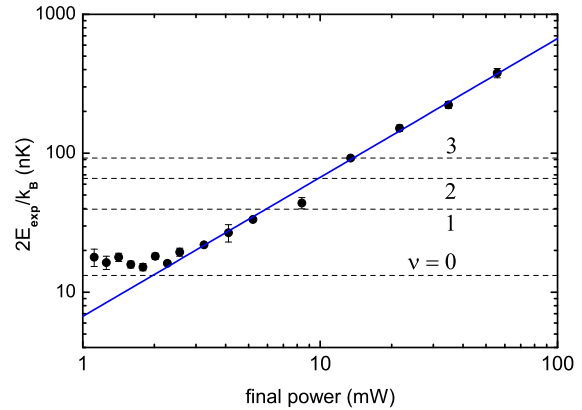}
\caption{\label{2Dtemperature} Vertical expansion energy versus
final power $P_{\rm f}$ of the evaporation ramp. The solid line
illustrates the proportionality to the horizontal trap depth above
3\,mW. The dashed lines indicate the expansion energies for the
four lowest vibrational levels.}
\end{figure}

The measured vertical expansion energy is shown as a function of
the final ramp power $P_{\rm f}$ in Fig.~\ref{2Dtemperature}. With
decreasing power down to about 3\,mW, we observe that $E_{\rm
exp}$ linearly follows the reduced horizontal trap depth $\hat{U}$
(see solid line). This shows that evaporative cooling takes place
with an approximately fixed relation between temperature $T$ and
trap depth $\hat{U}$
($k_BT/\hat{U} \approx 6$). 
For power values below 3\,mW we observe that $E_{\rm exp}$ levels
off at about $\frac{1}{2}k_B \times 16\,$nK. 
This value is well below the vibrational energy quantum $\hbar
\omega = k_B \times 26\,$nK and close to the zero point energy of
$k_B \times 13$\,nK. A comparison with the energies $E_\nu = (\nu
+ \frac{1}{2})\hbar\omega_z$ of the quantum states in the vertical
motion (see dashed lines in Fig.~\ref{2Dtemperature}) shows that a
two-dimensional gas is realized. This atomic ``pancake'' is
created in a surface trap with an aspect ratio of about 50:1.

According to a calculation of the phase-space density based on the
measured temperatures and the known trap frequencies, we expect
the onset of BEC approximately at a final ramp power $P_{\rm
f}\approx$10\,mW. Unfortunately, our release-and-recapture method
does not provide sufficient information to extract two-component
distributions in that regime and the data of
Fig.~\ref{2Dtemperature} do not show any signature of a phase
transition. We have therefore developed two other methods to show
the presence of a BEC in the surface trap.

Our first method to prove BEC relies on a controlled collapse of
the condensate at negative scattering length \cite{Roberts2001a},
a phenomenon that does not occur in a thermal gas. The scattering
length is known to be negative for magnetic fields below 17\,G and
the collapse of a Cs BEC below that field has been demonstrated in
\cite{Weber2003a}. In a series of collapse measurements performed
with a trapped sample at $P_{\rm f} =1.6$\,mW we switched the
magnetic field from the evaporation field of 26.8\,G to a variable
field between 0\,G ($a\approx-3000\,a_0$) and 40\,G ($a\approx
1000\,a_0$) for a short time interval of 20\,ms and measured the
resulting loss. When the scattering length is negative the
contraction of the BEC leads to a dramatic increase in density. In
this case three-body recombination causes very fast loss. In a
thermal gas, no contraction can happen and three-body loss at
moderate scattering length is too slow to cause any significant
reduction of the trapped atom number.


\begin{figure}
\includegraphics[width=6.5cm]{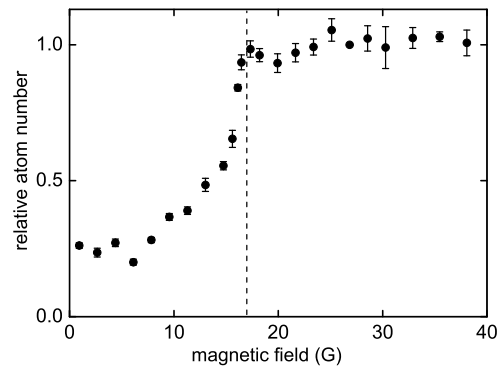}
\caption{\label{collapse}Atom loss showing the magnetically
induced BEC collapse at $P_{\rm f}= 1.6$mW. The magnetic field is
switched to a variable value for a short time interval 20\,ms. The
dashed line indicates the zero crossing of the scattering length
below which the self-interaction is attractive.}
\end{figure}

The magnetically induced collapse is demonstrated by the data in
Fig.~\ref{collapse}. For fields above the zero crossing at 17\,G
switching at $a>0$ does not induce any significant loss. Below
17\,G where $a<0$ a rapid decrease in the trapped atom number is
observed. Already at $a = -150\,a_0$ realized at 14.7\,G almost
half of the atoms are lost. This rapid loss at a field where the
three-body loss coefficient is relatively small can only be
explained by the huge density increase occurring during the
condensate collapse.

\begin{figure}
\includegraphics[width=6.5cm]{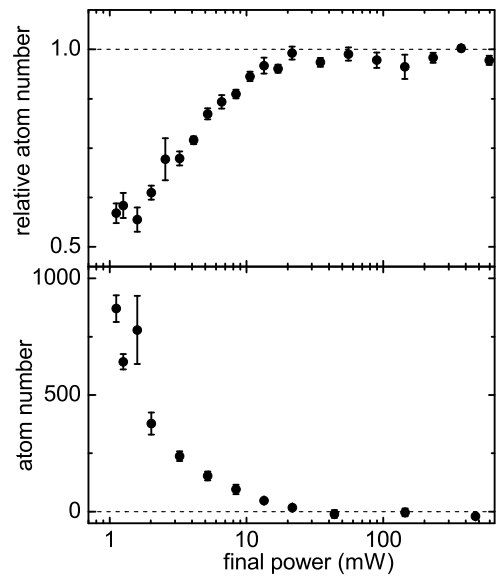}
\caption{\label{phase-transition} Observation of the phase
transition by two different methods. (a) The fraction of atoms
remaining after a magnetically induced condensate collapse at
$a=-150a_0$ is plotted as a function of the final ramp power
$P_{\rm f}$. (b) The horizontal release-and-recapture measurements
show the remaining number of atoms after 400\,ms of horizontal
expansion.}
\end{figure}

In order to identify the final evaporation ramp power $P_{\rm f}$
where the phase transition occurs, we measure the induced collapse
loss as a function of $P_{\rm f}$. In
Fig.~\ref{phase-transition}(a) we plot the fraction of atoms that
is observed to remain after a 20-ms application of a collapse
field of 14.7\,G. Above $P_{\rm f}\approx20$\,mW no significant
loss is observed as expected for a thermal gas, which cannot
undergo a collapse. Below 20\,mW rapidly increasing loss sets in,
which we interpret as the collapse of a growing condensate
fraction. From this observation we infer that the phase transition
takes place at a power of $P_{\rm f} \approx 20$\,mW, where we
have 15\,000 atoms left at a temperature of 150\,nK.
This is somewhat above the ramp power of
$\sim$10\,mW that we would expect from calculations of the
phase-space density, but still lies in the corresponding error
range.

Our second method to observe the onset of BEC is based on a
horizontal release-and-recapture technique. After evaporation with
a final ramp power $P_{\rm f}$ we turn off the horizontally
confining attractive laser beam for a long time interval of
400\,ms. For recapturing the remaining atoms we then turn on the
horizontally confining beam at a fixed power of 22\,mW. No atoms
are recaptured for thermal samples ($P_{\rm f}
> 20$\,mW) as shown in
Fig.~\ref{phase-transition}(b). For deeper evaporation with
$P_{\rm f}$ below 20\,mW we observe a rapidly increasing number of
recaptured atoms which we interpret as a consequence of the much
slower expansion of a condensate. The fact that the recapture
signal occurs at the same final power as the onset of the collapse
loss strongly supports this interpretation.

In the horizontal expansion measurements we observe significant
recapture signals even after a few seconds of free evolution on
the EW without the horizontally confining beam. This can not be
explained by a slow expansion alone but rather indicates the
presence of a horizontal trapping mechanism. An explanation might
be the roughness of an EW atom mirror \cite{Henkel1997a} leading
to shallow potential wells. Surface imperfections of the prism or
the interference with residual stray light can produce
corrugations in the optical potential of the evanescent wave. A
trapping effect could thus emerge from the pinning of the
condensate to such defects.

With better knowledge on the phase transition we finally revisit
the data of Fig.~\ref{2Dtemperature}. The transition takes place
at $P_{\rm f} \approx 20$\,mW in a 3D regime when the thermal
component populates about five vertical quantum states. With
decreasing values of $P_{\rm f}$ a crossover to a 2D regime takes
place. For $P_{\rm f} \lesssim 2$\,mW a 2D BEC is formed, for
which we estimate a chemical potential of $k_B \times 5$nK. In
this 2D regime, the zero-point energy of $k_B \times 13$nK
dominates the vertical motion.

In order to further explore the properties of the 2D BEC in our
GOST we are presently developing a new apparatus that offers much
faster magnetic field control and much better optical access.
These improvements will enable us to implement both in-situ and
expansion imaging techniques. Magnetic levitation will allow us to
overcompensate gravity and thus to observe a freely expanding atom
cloud after release from the surface trap. For enhancing the
anisotropy the two options are magnetic compression against the EW
or the
application of an additional attractive EW \cite{Hammes2003a}. 
The trapped two-dimensional condensate will allows us to study
elementary excitations like solitons and vortices, the properties
of which may exhibit striking differences as compared to the
three-dimensional case. Further intriguing possibilities are
offered by the creation of optical surface lattices created
through the interference of different evanescent waves.

In conclusion, we have realized BEC in an optical surface trap
based on an evanescent-wave atom mirror. During evaporative
cooling we reach the phase transition in a 3D situation before the
condensate is brought into the 2D regime by further increasing the
trap anisotropy. The system opens up new possibilities to study
the widely unexplored properties of degenerate quantum gases in
two dimensions.

\begin{acknowledgments}
We gratefully acknowledge support by the Austrian Science Fund
(FWF) within SFB 15 (project part 15) and by the European
Community through the Research Training Network ``FASTNet'' under
contract No.\ HPRN-CT-2002-00304.

\end{acknowledgments}


\begin{thebibliography}{21}
\expandafter\ifx\csname
natexlab\endcsname\relax\def\natexlab#1{#1}\fi
\expandafter\ifx\csname bibnamefont\endcsname\relax
  \def\bibnamefont#1{#1}\fi
\expandafter\ifx\csname bibfnamefont\endcsname\relax
  \def\bibfnamefont#1{#1}\fi
\expandafter\ifx\csname citenamefont\endcsname\relax
  \def\citenamefont#1{#1}\fi
\expandafter\ifx\csname url\endcsname\relax
  \def\url#1{\texttt{#1}}\fi
\expandafter\ifx\csname
urlprefix\endcsname\relax\def\urlprefix{URL }\fi
\providecommand{\bibinfo}[2]{#2}
\providecommand{\eprint}[2][]{\url{#2}}

\bibitem[{\citenamefont{Pricoupenko et~al.}(2003)\citenamefont{Pricoupenko,
  Perrin, and Olshanii}}]{QGLD2003}
\bibinfo{editor}{\bibfnamefont{L.}~\bibnamefont{Pricoupenko}},
  \bibinfo{editor}{\bibfnamefont{H.}~\bibnamefont{Perrin}}, \bibnamefont{and}
  \bibinfo{editor}{\bibfnamefont{M.}~\bibnamefont{Olshanii}}, eds.,
  \emph{\bibinfo{title}{Proceedings of the Euroschool on Quantum Gases in Lower
  Dimensions, Les Houches, 15 - 25 April 2003}} (\bibinfo{year}{2003}),
  \bibinfo{note}{{J.} Phys. IV, in press}.

\bibitem[{\citenamefont{Bagnato and Kleppner}(1991)}]{Bagnato1991a}
\bibinfo{author}{\bibfnamefont{V.}~\bibnamefont{Bagnato}} \bibnamefont{and}
  \bibinfo{author}{\bibfnamefont{D.}~\bibnamefont{Kleppner}},
  \bibinfo{journal}{Phys. Rev. A} \textbf{\bibinfo{volume}{44}},
  \bibinfo{pages}{7439} (\bibinfo{year}{1991}).

\bibitem[{\citenamefont{Ketterle and van Druten}(1996)}]{Ketterle1996a}
\bibinfo{author}{\bibfnamefont{W.}~\bibnamefont{Ketterle}} \bibnamefont{and}
  \bibinfo{author}{\bibfnamefont{N.~J.} \bibnamefont{van Druten}},
  \bibinfo{journal}{Phys. Rev. A} \textbf{\bibinfo{volume}{54}},
  \bibinfo{pages}{656} (\bibinfo{year}{1996}).

\bibitem[{\citenamefont{G{\"orlitz} et~al.}(2001)\citenamefont{G{\"orlitz},
  Vogels, Leanhardt, Raman, Gustavson, Abo-Shaeer, Chikkatur, Gupta, Inouye,
  Rosenband et~al.}}]{Gorlitz2001a}
\bibinfo{author}{\bibfnamefont{A.}~\bibnamefont{G{\"orlitz}}},
  \bibinfo{author}{\bibfnamefont{J.}~\bibnamefont{Vogels}},
  \bibinfo{author}{\bibfnamefont{A.}~\bibnamefont{Leanhardt}},
  \bibinfo{author}{\bibfnamefont{C.}~\bibnamefont{Raman}},
  \bibinfo{author}{\bibfnamefont{T.}~\bibnamefont{Gustavson}},
  \bibinfo{author}{\bibfnamefont{J.}~\bibnamefont{Abo-Shaeer}},
  \bibinfo{author}{\bibfnamefont{A.}~\bibnamefont{Chikkatur}},
  \bibinfo{author}{\bibfnamefont{S.}~\bibnamefont{Gupta}},
  \bibinfo{author}{\bibfnamefont{S.}~\bibnamefont{Inouye}},
  \bibinfo{author}{\bibfnamefont{T.}~\bibnamefont{Rosenband}},
  \bibnamefont{et~al.}, \bibinfo{journal}{Phys. Rev. Lett.}
  \textbf{\bibinfo{volume}{87}}, \bibinfo{pages}{130402}
  (\bibinfo{year}{2001}).

\bibitem[{\citenamefont{Schreck et~al.}(2001)\citenamefont{Schreck, Khaykovich,
  Corwin, Ferrari, Bourdel, Cubizolles, and Salomon}}]{Schreck2001a}
\bibinfo{author}{\bibfnamefont{F.}~\bibnamefont{Schreck}},
  \bibinfo{author}{\bibfnamefont{L.}~\bibnamefont{Khaykovich}},
  \bibinfo{author}{\bibfnamefont{K.~L.} \bibnamefont{Corwin}},
  \bibinfo{author}{\bibfnamefont{G.}~\bibnamefont{Ferrari}},
  \bibinfo{author}{\bibfnamefont{T.}~\bibnamefont{Bourdel}},
  \bibinfo{author}{\bibfnamefont{J.}~\bibnamefont{Cubizolles}},
  \bibnamefont{and} \bibinfo{author}{\bibfnamefont{C.}~\bibnamefont{Salomon}},
  \bibinfo{journal}{Phys. Rev. Lett.} \textbf{\bibinfo{volume}{87}},
  \bibinfo{pages}{080403} (\bibinfo{year}{2001}).

\bibitem[{\citenamefont{Walraven}()}]{Walraven1991a}
\bibinfo{author}{\bibfnamefont{J.~T.~M.} \bibnamefont{Walraven}},
  \bibinfo{howpublished}{in {\it Fundamental Systems in Quantum Optics}, edited
  by J. Dalibard, J.M. Raimond, and J. Zinn-Justin (Elsevier, Amsterdam, 1992),
  p. 485.}

\bibitem[{\citenamefont{Safonov et~al.}(1998)\citenamefont{Safonov, Vasilyev,
  Yasnikov, Lukashevich, and Jaakkola}}]{Safonov1998a}
\bibinfo{author}{\bibfnamefont{A.~I.} \bibnamefont{Safonov}},
  \bibinfo{author}{\bibfnamefont{S.~A.} \bibnamefont{Vasilyev}},
  \bibinfo{author}{\bibfnamefont{I.~S.} \bibnamefont{Yasnikov}},
  \bibinfo{author}{\bibfnamefont{I.~I.} \bibnamefont{Lukashevich}},
  \bibnamefont{and} \bibinfo{author}{\bibfnamefont{S.}~\bibnamefont{Jaakkola}},
  \bibinfo{journal}{Phys. Rev. Lett.} \textbf{\bibinfo{volume}{81}},
  \bibinfo{pages}{4545} (\bibinfo{year}{1998}).

\bibitem[{\citenamefont{Kagan et~al.}(1987)\citenamefont{Kagan, Svistunov, and
  Shlyapnikov}}]{Kagan1987a}
\bibinfo{author}{\bibfnamefont{Y.}~\bibnamefont{Kagan}},
  \bibinfo{author}{\bibfnamefont{B.}~\bibnamefont{Svistunov}},
  \bibnamefont{and}
  \bibinfo{author}{\bibfnamefont{G.}~\bibnamefont{Shlyapnikov}},
  \bibinfo{journal}{Sov. Phys. JETP} \textbf{\bibinfo{volume}{66}},
  \bibinfo{pages}{314} (\bibinfo{year}{1987}).

\bibitem[{\citenamefont{Petrov et~al.}(2000)\citenamefont{Petrov, Holzmann, and
  Shlyapnikov}}]{Petrov2000a}
\bibinfo{author}{\bibfnamefont{D.}~\bibnamefont{Petrov}},
  \bibinfo{author}{\bibfnamefont{M.}~\bibnamefont{Holzmann}}, \bibnamefont{and}
  \bibinfo{author}{\bibfnamefont{G.}~\bibnamefont{Shlyapnikov}},
  \bibinfo{journal}{Phys. Rev. Lett.} \textbf{\bibinfo{volume}{84}},
  \bibinfo{pages}{2551} (\bibinfo{year}{2000}).

\bibitem[{\citenamefont{Grimm et~al.}(2000)\citenamefont{Grimm,
  Weidem{\"u}ller, and Ovchinnikov}}]{Grimm2000a}
\bibinfo{author}{\bibfnamefont{R.}~\bibnamefont{Grimm}},
  \bibinfo{author}{\bibfnamefont{M.}~\bibnamefont{Weidem{\"u}ller}},
  \bibnamefont{and}
  \bibinfo{author}{\bibfnamefont{Y.}~\bibnamefont{Ovchinnikov}},
  \bibinfo{journal}{Adv. At. Mol. Opt. Phys.} \textbf{\bibinfo{volume}{42}},
  \bibinfo{pages}{95} (\bibinfo{year}{2000}).

\bibitem[{\citenamefont{Wallis et~al.}(1992)\citenamefont{Wallis, Dalibard, and
  Cohen-Tannoudji}}]{Wallis1992a}
\bibinfo{author}{\bibfnamefont{H.}~\bibnamefont{Wallis}},
  \bibinfo{author}{\bibfnamefont{J.}~\bibnamefont{Dalibard}}, \bibnamefont{and}
  \bibinfo{author}{\bibfnamefont{C.}~\bibnamefont{Cohen-Tannoudji}},
  \bibinfo{journal}{Appl. Phys. B} \textbf{\bibinfo{volume}{54}},
  \bibinfo{pages}{407} (\bibinfo{year}{1992}).

\bibitem[{\citenamefont{Wallis}(1996)}]{Wallis1996a}
\bibinfo{author}{\bibfnamefont{H.}~\bibnamefont{Wallis}},
  \bibinfo{journal}{Quantum Semiclass. Opt.} \textbf{\bibinfo{volume}{8}},
  \bibinfo{pages}{727} (\bibinfo{year}{1996}).

\bibitem[{\citenamefont{Ovchinnikov et~al.}(1997)\citenamefont{Ovchinnikov,
  Manek, and Grimm}}]{Ovchinnikov1997a}
\bibinfo{author}{\bibfnamefont{Y.~B.} \bibnamefont{Ovchinnikov}},
  \bibinfo{author}{\bibfnamefont{I.}~\bibnamefont{Manek}}, \bibnamefont{and}
  \bibinfo{author}{\bibfnamefont{R.}~\bibnamefont{Grimm}},
  \bibinfo{journal}{Phys. Rev. Lett.} \textbf{\bibinfo{volume}{79}},
  \bibinfo{pages}{2225} (\bibinfo{year}{1997}).

\bibitem[{\citenamefont{Hammes et~al.}(2002)\citenamefont{Hammes, Rychtarik,
  N{\"a}gerl, and Grimm}}]{Hammes2002a}
\bibinfo{author}{\bibfnamefont{M.}~\bibnamefont{Hammes}},
  \bibinfo{author}{\bibfnamefont{D.}~\bibnamefont{Rychtarik}},
  \bibinfo{author}{\bibfnamefont{H.-C.} \bibnamefont{N{\"a}gerl}},
  \bibnamefont{and} \bibinfo{author}{\bibfnamefont{R.}~\bibnamefont{Grimm}},
  \bibinfo{journal}{Phys. Rev. A} \textbf{\bibinfo{volume}{66}},
  \bibinfo{pages}{051401(R)} (\bibinfo{year}{2002}).

\bibitem[{tra()}]{trapfreq}
\bibinfo{note}{The loss maximum occurs somewhat below the trap frequency
  because of the anharmonicity of the potential.}

\bibitem[{\citenamefont{Kerman et~al.}(2001)\citenamefont{Kerman, Chin,
  Vuleti{\'c}, Chu, Leo, Williams, and Julienne}}]{Kerman2001a}
\bibinfo{author}{\bibfnamefont{A.~J.} \bibnamefont{Kerman}},
  \bibinfo{author}{\bibfnamefont{C.}~\bibnamefont{Chin}},
  \bibinfo{author}{\bibfnamefont{V.}~\bibnamefont{Vuleti{\'c}}},
  \bibinfo{author}{\bibfnamefont{S.}~\bibnamefont{Chu}},
  \bibinfo{author}{\bibfnamefont{P.~J.} \bibnamefont{Leo}},
  \bibinfo{author}{\bibfnamefont{C.~J.} \bibnamefont{Williams}},
  \bibnamefont{and} \bibinfo{author}{\bibfnamefont{P.~S.}
  \bibnamefont{Julienne}}, \bibinfo{journal}{C. R. Acad. Sci. Paris IV}
  \textbf{\bibinfo{volume}{2}}, \bibinfo{pages}{633} (\bibinfo{year}{2001}).

\bibitem[{\citenamefont{Weber et~al.}(2003{\natexlab{a}})\citenamefont{Weber,
  Herbig, Mark, N{\"a}gerl, and Grimm}}]{Weber2003a}
\bibinfo{author}{\bibfnamefont{T.}~\bibnamefont{Weber}},
  \bibinfo{author}{\bibfnamefont{J.}~\bibnamefont{Herbig}},
  \bibinfo{author}{\bibfnamefont{M.}~\bibnamefont{Mark}},
  \bibinfo{author}{\bibfnamefont{H.-C.} \bibnamefont{N{\"a}gerl}},
  \bibnamefont{and} \bibinfo{author}{\bibfnamefont{R.}~\bibnamefont{Grimm}},
  \bibinfo{journal}{Science} \textbf{\bibinfo{volume}{299}},
  \bibinfo{pages}{232} (\bibinfo{year}{2003}{\natexlab{a}}).

\bibitem[{\citenamefont{Weber et~al.}(2003{\natexlab{b}})\citenamefont{Weber,
  Herbig, Mark, N{\"a}gerl, and Grimm}}]{Weber2003b}
\bibinfo{author}{\bibfnamefont{T.}~\bibnamefont{Weber}},
  \bibinfo{author}{\bibfnamefont{J.}~\bibnamefont{Herbig}},
  \bibinfo{author}{\bibfnamefont{M.}~\bibnamefont{Mark}},
  \bibinfo{author}{\bibfnamefont{H.-C.} \bibnamefont{N{\"a}gerl}},
  \bibnamefont{and} \bibinfo{author}{\bibfnamefont{R.}~\bibnamefont{Grimm}},
  \bibinfo{journal}{Phys. Rev. Lett.} \textbf{\bibinfo{volume}{91}},
  \bibinfo{pages}{123201} (\bibinfo{year}{2003}{\natexlab{b}}).

\bibitem[{\citenamefont{Hammes et~al.}(2003)\citenamefont{Hammes, Rychtarik,
  Engeser, N{\"a}gerl, and Grimm}}]{Hammes2003a}
\bibinfo{author}{\bibfnamefont{M.}~\bibnamefont{Hammes}},
  \bibinfo{author}{\bibfnamefont{D.}~\bibnamefont{Rychtarik}},
  \bibinfo{author}{\bibfnamefont{B.}~\bibnamefont{Engeser}},
  \bibinfo{author}{\bibfnamefont{H.-C.} \bibnamefont{N{\"a}gerl}},
  \bibnamefont{and} \bibinfo{author}{\bibfnamefont{R.}~\bibnamefont{Grimm}},
  \bibinfo{journal}{Phys. Rev. Lett.} \textbf{\bibinfo{volume}{90}},
  \bibinfo{pages}{173001} (\bibinfo{year}{2003}).

\bibitem[{\citenamefont{Roberts et~al.}(2001)\citenamefont{Roberts, Clausen,
  Cornish, Donley, Cornell, and Wieman}}]{Roberts2001a}
\bibinfo{author}{\bibfnamefont{J.~L.} \bibnamefont{Roberts}},
  \bibinfo{author}{\bibfnamefont{N.~R.} \bibnamefont{Clausen}},
  \bibinfo{author}{\bibfnamefont{S.~L.} \bibnamefont{Cornish}},
  \bibinfo{author}{\bibfnamefont{E.~A.} \bibnamefont{Donley}},
  \bibinfo{author}{\bibfnamefont{E.~A.} \bibnamefont{Cornell}},
  \bibnamefont{and} \bibinfo{author}{\bibfnamefont{C.~E.}
  \bibnamefont{Wieman}}, \bibinfo{journal}{Phys. Rev. Lett.}
  \textbf{\bibinfo{volume}{86}}, \bibinfo{pages}{4211} (\bibinfo{year}{2001}).

\bibitem[{\citenamefont{Henkel et~al.}(1997)\citenamefont{Henkel, Molmer,
  Kaiser, Vansteenkiste, Westbrook, and Aspect}}]{Henkel1997a}
\bibinfo{author}{\bibfnamefont{C.}~\bibnamefont{Henkel}},
  \bibinfo{author}{\bibfnamefont{K.}~\bibnamefont{Molmer}},
  \bibinfo{author}{\bibfnamefont{R.}~\bibnamefont{Kaiser}},
  \bibinfo{author}{\bibfnamefont{N.}~\bibnamefont{Vansteenkiste}},
  \bibinfo{author}{\bibfnamefont{C.~I.} \bibnamefont{Westbrook}},
  \bibnamefont{and} \bibinfo{author}{\bibfnamefont{A.}~\bibnamefont{Aspect}},
  \bibinfo{journal}{Phys. Rev. A} \textbf{\bibinfo{volume}{55}},
  \bibinfo{pages}{1160} (\bibinfo{year}{1997}).

\end{thebibliography}

\end{document}